# Data Colonisation and AI Governance in South Africa: Data Protection, Algorithmic Power, and the Struggle for Digital Sovereignty

**Takudzwa Musekiwa[1,2] · Kimon Kieslich[3,4]**

## Abstract

Artificial intelligence (AI) is not only a technological artefact; it is also a new site of political economy, legal struggle, and sovereignty contestation. AI systems depend on the continuous collection, storage, processing, and monetisation of large volumes of data. In South Africa, these datasets increasingly include behavioural, biometric, financial, consumer, location, health, labour, and public-sector information. Our article uses data colonisation as a theoretical lens to analyse whether South Africa's existing legal and governance framework can regulate AI-specific risks. It argues that South Africa's current laws, including the Protection of Personal Information Act (POPIA), regulate personal information but do not adequately prevent data colonisation. POPIA remains important for privacy, lawful processing, automated decision-making, and cross-border transfers. However, data colonisation cannot solely be reduced to privacy protection. It concerns ownership, dependency, infrastructure, surveillance, regulatory power, value extraction, and the capacity of people and institutions to challenge algorithmic decisions. The article examines South African AI-related risks in the areas of cloud and digital infrastructure, biometric surveillance, platform work, financial profiling, digital identity, public-sector automation, and digital markets. This paper argues that South Africa's legal framework does not yet adequately regulate algorithmic opacity, externally controlled digital infrastructure, biometric surveillance, automated profiling, collective data harms, or the unequal extraction of value from South African data. We argue for the need of a decolonial AI governance framework based on algorithmic impact assessments, public-sector AI registers, independent audits, biometric safeguards, competition-law remedies, public-interest data governance, stronger regulatory capacity, and democratic control to strengthen South Africa's digital future.

---

✉ tamusekiwa@gmail.com, k.kieslich@uva.nl

[1] Southern Centre for Inequality Studies (SCIS), University of Witwatersrand, South Africa

[2] Centre for Advancement of Scholarship, University of Pretoria, South Africa

[3] Amsterdam School of Communication Research, University of Amsterdam, The Netherlands

[4] African German Centre for Sustainable and Resilient Food Systems and Applied Agricultural and Food Data Science (UKUDLA), University of Hohenheim, Germany

## 1. Introduction

Data colonisation should be the starting point for any serious discussion of artificial intelligence (AI) governance in South Africa. The central issue is not merely that AI systems process personal information or that privacy may be infringed. The deeper issue is that the infrastructures and decision-making systems through which data becomes social and economic power are often controlled by actors whose interests are not democratically accountable to people in South Africa [14, 19, 64]. AI therefore forces South Africa to confront a foundational set of questions: Who collects data? Who owns the data? Where is data stored? Who builds the AI systems that process it? Who profits from the data? Who is harmed when algorithmic systems make biased or opaque decisions? And who has the legal, technical, financial, and political power to challenge those decisions? These are not simply questions of privacy. They are questions of ownership, dependency, surveillance, sovereignty, and regulatory power.

The importance of these questions becomes clearer when AI is understood as a data-intensive mode of production. Machine-learning systems do not become powerful simply because of code. They become powerful because they are trained, optimised, tested, and refined on large volumes of data [24]. The OECD notes that building machine-learning models often requires large volumes and varieties of data, including data that includes personal information [50]. Large language models and other foundation models have intensified this dependence by drawing on web-scale datasets, digitised documents, code repositories, images, voice recordings, platform interactions, and user-generated content [23, 66]. As Dodge et al. show in their study of the Colossal Clean Crawled Corpus, large web-text datasets are often created from massive internet crawls with limited documentation, uneven representation, and filtering choices that can remove or underrepresent minority groups [30]. In AI governance, therefore, data is not merely an input. It is the raw material of algorithmic power.

For Africa, and for South Africa in particular, this creates a risk that the information age reproduces older forms of colonial extraction in new technological forms. This article therefore uses the term data colonisation to describe an asymmetrical relationship in which data generated by people, communities, institutions, and territories is extracted, stored, processed, analysed, and monetised in ways that concentrate power elsewhere. Segev's work on information and power is useful here because it reminds us that information is not neutral; it becomes powerful when certain actors are able to control the stories, classifications, and knowledge systems through which social life is interpreted [66]. More directly in terms of data colonialism, Couldry and Mejias argue that contemporary data relations appropriate human life itself as a raw material for capitalist accumulation [22, 23]. Birhane similarly describes algorithmic colonisation as a process through which AI systems developed in powerful technological centres are deployed in African contexts without adequate attention to local values, histories, inequalities, and epistemologies [14]. Kwet extends this argument by linking digital colonialism to the dominance of Global North corporations over hardware, software, cloud infrastructure, intellectual property, and digital platforms in the Global South [40].

Our focus on South Africa is deliberate. South Africa is one of the most digitally advanced economies on the continent. UNESCO finds that South Africa combines relatively strong AI governance and digital infrastructure with persistent gaps in computing capacity, data-sharing frameworks, digital skills, and inclusive participation [70]. South Africa has a constitutional right to privacy, a comprehensive data-protection statute in the Protection of Personal Information Act (POPIA), an Information Regulator, sophisticated financial and telecommunications sectors, rapidly expanding cloud and data-centre markets, and active debate about AI policy. However, South Africa is also a deeply unequal society marked by apartheid's spatial, racial, economic, and administrative legacies. AI systems operating in South Africa therefore enter a challenging social environment. These systems operate in a society shaped by apartheid where race, class, geography, language, disability, gender, and citizenship status continue to structure access to services, employment, credit, mobility, and state protection. In South Africa, algorithmic systems trained on biased data or deployed without effective oversight can reproduce these inequalities while appearing objective, technical, or inevitable.

It is also important to point out that South Africa's current legal framework has some safeguards in place that are aiming to govern digital technologies. POPIA regulates personal information and establishes conditions for lawful processing (POPI Act, Republic of South Africa 2020). Section 71 of the Act regulates certain forms of solely automated decision-making that result in legal consequences or substantially affect a data subject. Section 72 regulates cross-border transfers of personal information. South Africa's Consumer protection law [59] can address unfair marketing and consumer harm. Competition law [55] can address platform dominance and anti-competitive conduct in digital markets. Financial regulators have begun examining AI use in the financial sector [31]. The Department of Communications and Digital Technologies has produced an AI Policy Framework and attempts to develop a national AI policy [27]. The African Union's Data Policy Framework and Continental AI Strategy also provide regional guidance. These instruments matter, and we will refer to the existing body of regulatory instruments throughout the article [4, 5].

Our central argument is, however, that South Africa's current policy frameworks and laws do not adequately regulate AI-specific risks such as algorithmic opacity, foreign-controlled digital infrastructure, biometric surveillance, automated profiling, and the unequal extraction of value from South African data. In reviewing the academic literature on academic literature on data colonialism, algorithmic governance and AI ethics and existing policies and regulations, the article examines whether South Africa's legal and governance architecture is capable of responding to AI-driven data colonisation.

The article proceeds in five steps. First, it discusses the literature on data colonisation as the theoretical lens, linking digital extraction to historical colonialism and defining the indicators through which data colonisation can be identified in South Africa. Second, it examines how data colonisation happens through AI systems in South Africa, focusing on data extraction, infrastructure dependency, value extraction, biometric surveillance, automated profiling, algorithmic bias, and limited redress. Third, it analyses South Africa's legal and governance framework, including POPIA, constitutional rights, consumer and competition law, financial regulation, public-sector governance, and the evolving AI policy process. Fourth, it explains why existing laws do not (fully) prevent data colonisation. Finally, it proposes a decolonial AI governance agenda rooted in digital sovereignty, algorithmic accountability, democratic oversight, and public-interest data governance.

## 2. Data Colonisation

Data colonisation serves as the conceptual lens for this article. The term is used to describe a new form of extraction that operates through datafication, digital platforms, AI systems, cloud infrastructure, and algorithmic decision-making. Couldy and Mejias define data colonialism as a process through which human life is appropriated through data relations for capitalist accumulation [23]. Their argument is not merely that companies collect too much data. It is that everyday life is being reorganised so that social relations, movements, preferences, expressions, labour, transactions, and interactions become raw material for extraction. This is why data colonialism goes beyond surveillance and privacy protection.

Segev's work on information and power deepen this theoretical argument [66]. Segev argues that information becomes powerful when it can define what counts as knowledge, whose behaviour is visible, whose risk is calculable, whose identity is verified, whose creditworthiness is recognised, and whose claims are ignored [66]. In the context of AI, information power becomes embedded in models, scoring systems, recommender systems, databases, and automated decisions. Data colonisation can therefore be understood as the asymmetrical extraction and conversion of information into decision-making power.

Consequently, this article defines data colonisation as follows: Data colonisation is the asymmetrical extraction, storage, processing, ownership, and monetisation of data from people, communities, institutions, and territories in ways that concentrate infrastructural, economic, epistemic, and regulatory power in the hands of dominant external or private actors, while limiting the ability of those who generate the data to control, benefit from, audit, contest, or govern its use.

The link to historical colonialism is not metaphorical only. Historical colonialism worked through a structure of extraction and control [19, 45]: Land was surveyed, mapped, enclosed, and transferred. Minerals were mined and exported. Labour was coerced, disciplined, and racialised [77]. Indigenous knowledge was appropriated or devalued. Colonial administrations produced records, censuses, pass systems, maps, and categories that made populations governable [77]. In South Africa, apartheid intensified this administrative logic through racial classification, spatial control, labour regulation, surveillance, and bureaucratic exclusion. Data colonisation does not repeat colonialism in identical form, but it carries forward the logic of extraction, classification, dependency, and external control.

The information age and the so-called Fourth Industrial Revolution have changed the object of extraction. Instead of land and minerals alone, the new resource is data: biometric data from faces, fingerprints, and identity systems; behavioural data from platforms and devices; financial data from transactions and credit scoring; health data from clinics, insurers, and digital health platforms; location data from mobile phones and vehicles; labour data from platform work and workplace monitoring; consumer data from e-commerce and advertising systems; and institutional data from public services, education, policing, transport, welfare, and local government [24, 67]. This does not mean land, minerals, and labour no longer matter. Rather, data becomes an additional layer through which existing inequalities are measured, managed, monetised, and reproduced.

Birhane warns that AI systems deployed in African contexts often carry assumptions, values, and design priorities from elsewhere and argues that those systems can become a form of colonisation when they impose external standards and extract local data without meaningful local accountability [14]. In the same vein, Segun argues that the corporate structure of AI adoption is in fundamental opposition to central Africa values such as communitarianism and human dignity [67]. Mohamed et al. state that, in consequence, *decolonial* AI must examine power asymmetries in the production, deployment, and evaluation of AI systems [45]. Kwet frames digital colonialism as a structural relationship in which corporations from dominant economies control the digital infrastructure of the Global South: hardware, software, cloud services, platforms, standards, and intellectual property [40]. These arguments are directly relevant to South Africa because the country already hosts one of Africa's most developed data-centre and cloud-services ecosystems, including hyperscale infrastructure operated by Microsoft, Amazon Web Services and Google [32, 70]. However, the localisation of physical infrastructure does not necessarily translate into technological sovereignty, since much of the hyperscale cloud infrastructure available in South Africa is provided by multinational technology corporations, while domestic high-performance computing capacity and data-sovereignty concerns remain significant [32, 70].

In platform environments, users or businesses may have little real choice because social, financial, labour, or public services are mediated through digital systems [43, 73]. The problem is compounded when data is used not only to provide a service but also to train models, build profiles, optimise advertising, improve surveillance, predict behaviour, or create commercial products [24]. Scholars have further argued that deploying AI models essentially makes users dependent on big tech cloud infrastructure, which is usually located at Big Tech companies from the US [72]. Hence, data can be lawfully collected under privacy law and still contribute to colonisation if the value and power generated from that data are captured elsewhere.

A useful way to identify data colonisation in South Africa is to look for indicators rather than isolated incidents. First, there is data extraction [19, 64, 74]: AI systems collect or depend on large volumes of data generated in South Africa, including personal, health, biometric, consumer, location, education, labour, and public-sector data. Second, there is infrastructure dependency [19, 64]: the data is stored, processed, analysed, or governed through foreign-owned platforms, cloud systems, hardware, or proprietary models. This is not inherently bad because global cloud services can improve efficiency, cybersecurity, scalability, and access to advanced tools. The problem arises when South African regulators, researchers, public bodies, workers, and communities cannot inspect, audit, control, or negotiate the terms of this dependency. Third, there is value extraction [14, 19, 63, 74]: South African data becomes a source of profit, market dominance, model improvement, or platform intelligence for powerful companies, while the local benefits are limited or uneven. Fourth, there is opacity [9, 74]: people affected by AI decisions, but also governments relying on AI technologies of the Global North can often not understand, challenge, or correct the models that affect them or are implemented in society, respectively. Fifth, there is surveillance and control [19, 64]: data is used to monitor bodies, movements, work performance, neighbourhoods, welfare access, or consumer behaviour. Sixth, there is weak redress [63, 74]: affected individuals or groups lack the legal, technical, and institutional means to challenge algorithmic decisions. Further, also governments in Africa themselves have few possibilities to counteract or influence AI development, which is hugely dominated by companies from the Global North.

These indicators mirror colonial extraction. In historical colonialism, local resources were extracted, processed elsewhere, transformed into wealth, and returned as instruments of dependency [45]. In data colonisation, South African data may be collected locally but processed through external infrastructures, converted into proprietary intelligence, and returned as products, services, scores, or surveillance systems that South Africans must accept without meaningful control. The issue is not whether all foreign technology is harmful. It is whether South Africa can exercise democratic, legal, technical, and economic agency over the systems that increasingly govern social life.

## 3. Data Colonisation through AI in South Africa

South Africa already shows the conditions through which data colonisation can take shape. The country has a large and growing digital economy, expanding mobile connectivity, a sophisticated financial sector, increasingly digitised public services, and major investments in cloud and data-centre infrastructure [6, 32, 70]. These developments are not inherently negative. Digital infrastructure and digitalisation can support economic growth, productivity, service delivery, financial and digital inclusion, research, and innovation [6, 78]. The concern is that the data generated by these systems may be governed by legal frameworks that were not designed for the scale, opacity, and cross-border nature of AI-driven data extraction. Table 1 summarizes the indicators for data colonization and provides examples how data colonization is already taking place in South Africa.

| Indicator | Meaning | South African examples |
|---|---|---|
| **Data extraction** | AI systems collect or depend on large volumes of South African behavioural, biometric, financial, health, labour, consumer, location, linguistic, and public-sector data. | Platform data, credit data, CCTV footage, biometric identity data, social grant data, workplace metrics, mobile and transaction data. |
| **Infrastructure dependency** | Data is stored, processed, analysed, or governed through foreign-owned or proprietary cloud, platform, model, hardware, or software infrastructures. | Cloud regions, hyperscaler platforms, proprietary AI models, surveillance vendors, platform APIs. |
| **Value extraction** | Economic value generated from South African data is captured by dominant firms, often with limited local ownership, benefit-sharing, taxation, or public-interest return. | Targeted advertising, model training, platform commissions, risk analytics, data brokerage, digital-market dominance. |
| **Algorithmic opacity** | Affected people cannot understand how AI systems classify, rank, score, include, exclude, approve, deny, price, or discipline them. | Credit scoring, insurance pricing, job allocation, platform deactivation, public-service eligibility, fraud detection. |
| **Surveillance and control** | Data is used to monitor bodies, workers, neighbourhoods, movement, identity, public space, or service access. | Smart CCTV, facial recognition, licence-plate recognition, biometric ID, employee monitoring, predictive policing. |
| **Weak redress** | Citizens, workers, consumers, and communities lack meaningful legal, technical, or institutional routes to challenge AI decisions or audit data use. | Opaque appeals, limited explanations, costly litigation, weak technical capacity in regulators, proprietary vendor secrecy. |

**Table 1 Indicators of data colonisation in South Africa**

The first mechanism is the mass *extraction of data*. AI systems in South Africa can draw on financial transactions, retail histories, mobile-network data, app usage, geolocation, biometric identifiers, workplace performance metrics, social media activity, call-centre records, school and university data, medical records, CCTV footage, and public-service interactions. This extraction is already visible in practice. In the financial sector, AI and machine-learning systems are used for purposes including credit scoring, fraud detection, transaction monitoring and customer profiling, turning financial and behavioural data into inputs for risk assessment and decision-making [31]. Similarly, South Africa's digital-identity ecosystem increasingly relies on biometric and administrative data to authenticate individuals and mediate access to services, illustrating how identity itself becomes a continually processed digital resource [51]. From a data-colonisation perspective, the concern is therefore not data collection alone, but the unequal power to control, combine, analyse and derive value from these datasets. Crucially, most AI services used in the African context are owned by Western companies. Hence, African countries are usually given up control over data to non-local companies, who are profiting from data from the African continent: "As external entities monetise these data, profits often flow out of Africa, contributing to a digital wealth transfer that exacerbates economic imbalances within the continent." [64:5]

The second mechanism is *infrastructure dependency*. South Africa hosts major cloud and data-centre infrastructure, but much of the hyperscale cloud ecosystem is owned by global firms. In South Africa data-centre and cloud investments are expanding by firms such as Amazon Web Services, Microsoft, Teraco, and Dimension Data [71]. Anwar notes that South Africa is the only African country where Amazon, Microsoft, and Google have built their own data centres [10]. This local presence may improve latency, reliability, and compliance with data-residency expectations. However, local data centres do not automatically create digital sovereignty. As Hawkins et al. argue in relation to AI compute sovereignty, the nationality of the companies that own and operate compute infrastructure matters [34]. A data centre located on South African soil may still be embedded in foreign corporate governance, foreign law, proprietary architectures, and global supply chains. The sovereignty question is therefore not only where data is stored, but who controls the infrastructure, software stack, audit logs, access rules, encryption architecture, and model deployment environment. Furthermore, a dominance of Western companies in the AI market stifles market entry possibilities for local companies as they need to compete for talent with companies who are more established and better funded [64]. In addition, the enormous technological affordances to handle large amounts of data gives (Western) Big Tech companies a structural advantage that is nearly impossible to bridge for local companies [19].

The third mechanism is *value extraction*. Data generated in South Africa can become a source of revenue for platforms, advertisers, financial institutions, AI vendors, data brokers, and cloud providers [19, 74]. This resembles colonial extraction when local resources generate external value while local communities receive limited benefit or bear disproportionate risk. In digital markets, the Competition Commission of South Africa has recognised that platform dominance can affect visibility, ranking, pricing, access, and market participation [20]. Its Online Intermediation Platforms Market Inquiry identified competition concerns in platform markets, and its later media and digital platforms work addressed the power of global platforms over news distribution and monetisation [20]. These competition developments are not framed as data colonisation, but they reveal a central point: control over digital intermediation allows powerful platforms to shape markets, extract rents, and determine which local actors are visible or viable. Coleman explains how these practices map onto the notion of digital colonialism: “[W]hen this business strategy is pursued in countries with limited infrastructure, limited data protection laws, and limited competition—while rooted in neoliberal notions of human rights—it transitions from a business model into a form of digital colonialism.” [19:428]

The fourth mechanism is *algorithmic opacity*, which is evident in *automated profiling* and financial *decision-making*. South Africa's financial sector is a leading site of AI adoption. The Prudential Authority and Financial Sector Conduct Authority's report on AI in the South African financial sector provides an important overview of current and planned uses of machine learning and generative AI [31]. AI can support fraud detection, anti-money laundering, customer service, risk management, and financial inclusion. But AI can also produce new forms of exclusion when credit, insurance, fraud, and affordability decisions are based on opaque models. Mota Makore argues that AI can support financial inclusion in South Africa by improving credit scoring and data analysis but also requires appropriate regulation [46]. The key problem is that a person denied credit, priced differently for insurance, flagged as risky, or excluded from a financial product may not know which data was used, how the model weighted it, or whether the result reflects historical inequality.

Algorithmic opacity is also evident in *platform work* and *algorithmic management* [45]. Location-based platform work in South Africa has grown through ride-hailing, delivery, courier services, domestic services, and other app-mediated work. The Labour Research Service notes that many such workers are treated as independent contractors and therefore lack the protections associated with standard employment. Algorithmic management complicates this further because work allocation, route planning, pay, ratings, incentives, penalties, and account access may be controlled through opaque platform systems. Anwar shows that ride-hailing drivers in Africa organise through self-organised networks to resist platform power [10], while Dinika highlights how algorithmic management shapes worker agency in Rwanda, South Africa, and Zimbabwe [28]. Data colonisation here is not only extraction of consumer data. It is extraction of worker data for managerial control, pricing, and platform profitability, with limited worker access to the data and logic that governs their livelihoods.

The fifth mechanism is *surveillance and control*, including *biometric surveillance*. South Africa's urban security environment has seen growing use of CCTV, licence-plate recognition, and data-driven surveillance systems. Vumacam and similar private surveillance infrastructures have been widely debated in Johannesburg and other urban areas [35]. From a data colonialism standpoint, those tools can lead to algorithmic oppression as most AI tools are not trained for the South African context and exert bias towards already marginalized population groups [45, 74]. Hunter describes the expansion of private surveillance in Johannesburg as a transformation of public space [35]. In practice, this can mean that ordinary movement through the city is converted into searchable data: licence-plate recognition systems can flag a vehicle as being of interest, prompting interception, investigation, detention, or vehicle impoundment. Such systems therefore do not merely observe public space; they can shape who is treated as suspicious and subjected to security or police intervention [35]. Scholars highlight the governance risks of facial recognition and digital surveillance in South Africa, including the risks of mass surveillance, function creep, weak transparency, and inadequate legal safeguards [36, 69]. Even where a system does not currently use facial recognition, the infrastructure may be capable of future integration with more intrusive analytics. Data colonisation appears here as the conversion of public movement into privately governed visual data, often without meaningful public debate. Surveillance and control are also evident through *digital identity* and public-sector data. Research ICT Africa's work on digital and biometric identity systems highlights the importance and risks of biometric identity in South Africa [51]. Razzano notes that public-sector biometric data collection in South Africa has long preceded modern data-protection provisions [51]. This is important because public-sector databases can become foundational infrastructures for welfare, identity, citizenship, mobility, fraud control, and service access. Razzano further examines exclusion risks in public digital systems and algorithmic decision-making [52]. The point is not that public-sector digitisation is inherently harmful. Rather, when biometric and administrative data become central to eligibility, verification, and service delivery, the governance framework must address due process, explainability, access, error correction, appeal, and collective harms. A privacy notice cannot remedy exclusion from a grant, identity service, or municipal process if the system's logic is opaque and the affected person cannot secure meaningful human review.

The sixth mechanism is *weak redress*, reflected in the *weak power to challenge* algorithmic decisions. Data colonisation is not only about extraction; it is also about asymmetry [63, 74]. A bank, platform, insurer, public agency, or surveillance provider may have legal teams, data scientists, proprietary models, cloud vendors, and contractual protections. A citizen, worker, grant applicant, migrant, informal trader, or low-income consumer may have none of these. They may not know that AI was used, which data shaped the decision, or may not be able to access the model, or may even not have the connectivity to meaningfully engage with AI systems. Further, they may not have resources to litigate, or face language, literacy, connectivity, disability, or bureaucratic barriers. The power to challenge is therefore central to the analysis. A legal right that cannot be practically exercised does not prevent data colonisation. Thus, decolonization practices are targeted towards giving people a voice who are currently not represented and giving them representation [74].

## 4. South Africa's Legal and Governance Framework

South Africa's legal and governance framework is more developed than that of many countries on the continent, but it remains fragmented [70]. The starting point is the Constitution [54]. Section 14 protects the right to privacy. Section 9 protects equality. Section 10 protects dignity. Section 16 protects freedom of expression. Section 32 protects access to information. Section 33 protects just administrative action. These rights are relevant to AI because algorithmic systems can affect dignity, equality, access to services, public participation, and administrative justice. The Constitution therefore provides a broad normative foundation for AI governance even in the absence of a dedicated AI statute.

POPIA is the main statute governing personal information. It gives effect to the constitutional right to privacy by regulating the processing of personal information by public and private bodies [60]. It establishes conditions for lawful processing, including accountability, processing limitation, purpose specification, further processing limitation, information quality, openness, security safeguards, and data-subject participation. It also creates rights and remedies for data subjects and establishes the Information Regulator as the key oversight body. POPIA matters because most AI systems that operate on identifiable personal data will fall within its scope. It is therefore the closest South African law currently comes to a cross-cutting AI governance instrument.

Section 71 of POPIA is particularly important because it addresses automated decision-making [70]. It provides that a data subject may not be subject to a decision that results in legal consequences or affects the person to a substantial degree if the decision is based solely on automated processing intended to provide a profile, including a profile related to work performance, creditworthiness, reliability, location, health, personal preferences, or conduct. This provision is directly relevant to AI systems used in credit scoring, recruitment, insurance, fraud detection, employment monitoring, and public-service eligibility. However, several scholars argue that section 71 leaves important uncertainties [25, 41, 49]. It focuses on decisions based solely on automated processing, which may allow organisations to avoid the provision by inserting nominal human involvement. It does not clearly provide a robust right to explanation. It does not address collective harms, model audits, or systemic bias. It does not create a general duty to conduct algorithmic impact assessments.

Section 72 of POPIA regulates cross-border transfers of personal information. It restricts transfers unless the recipient is subject to adequate protections, binding corporate rules, binding agreements, consent, contractual necessity, or other specified bases. This is important for cloud computing, platform services, global vendors, and AI model development. Yet section 72 is not a full data-sovereignty rule. It regulates transfer of personal information to third parties in foreign countries; it does not necessarily regulate model training on de-identified data, derived data, aggregated behavioural data, metadata, non-personal data, data used to improve proprietary systems, or the broader location and ownership of AI infrastructure. Scholars argue that cross-border data transfers under POPIA remain legally complex, especially where health, research, and data-sharing arrangements are involved [2, 42].

Consumer protection law also matters. The Consumer Protection Act also addresses unfair, unreasonable, or unjust contract terms, misleading marketing, unsafe goods and services, and unfair consumer practices [59]. In AI contexts, this may be relevant where automated systems shape marketing, pricing, eligibility, product recommendations, or service delivery. However, consumer law is not designed to audit AI systems. It does not create a general duty to disclose algorithmic profiling or explain model logic. It may only help *after* a harmful consumer practice occurs, but it is not a structural response to data colonisation.

Competition law has become increasingly important in the digital economy. The Competition Commission's digital markets work recognises that platform power can entrench dominance, affect discoverability, and limit opportunities for smaller firms. The Commission's Online Intermediation Platforms Market Inquiry and its media and digital platforms work show that South Africa can use competition law to address digital-market power [20]. The GovChat/Meta matter is also significant because it illustrates how a local civic-tech platform can become dependent on a global platform infrastructure. The Competition Commission alleged that Meta's attempt to off-board GovChat from WhatsApp raised abuse-of-dominance concerns, including GovChat's creation of large amounts of data that could support new services and products [20]. This is directly relevant to data colonisation: a domestic public-service platform may rely on a foreign platform's API, rules, and governance decisions even when it is serving South African public needs.

Financial regulation is another important source. The Prudential Authority and FSCA have started to examine AI use in the financial sector [31]. Their 2025 report is significant because it shows that AI adoption is already happening across the industry in banking, insurance, investments, payments, pensions, fintech, and lending. Financial-sector AI raises concerns about model governance, fairness, explainability, cybersecurity, outsourcing, cloud dependency, consumer disclosure, and operational risk. Sectoral regulators are therefore crucial to AI governance because many AI harms are domain specific. A general data-protection law cannot fully regulate technical risk in banking, insurance, credit, and fraud detection without sectoral expertise.

Cybersecurity and electronic communications law also form part of the broader framework. The Cybercrimes Act addresses unlawful access, interception and interference crime, and other related data and cyber offences [61]. The Electronic Communications and Transactions Act provides rules relevant to electronic transactions, consumer protection online, and automated digital processes [58]. These laws are important as AI systems depend on digital infrastructures and can create new cyber risks. However, they do not regulate whether AI systems are fair, explainable, auditable, locally accountable, or decolonial. They protect systems and transactions more than they regulate algorithmic power.

The national AI policy process remains unsettled. South Africa published a National AI Policy Framework in 2024 as a step towards a full national AI policy [27]. The Framework emphasised economic growth, innovation, human-centred AI, governance, ethics, talent, infrastructure, and inclusion. In April 2026, the government published a Draft National AI Policy, but it was withdrawn after fictitious or apparently AI-generated references were identified in the reference list [62]. As a consequence, an independent panel was formed to review the policy and the publication of a revised policy is planned for publication and public comment for January 2027 [29]. This incident is important for two reasons. First, South Africa does not yet have an enacted AI-specific law such as the EU AI Act. Second, the withdrawal itself illustrates why AI governance requires verified evidence, human oversight, transparency, and institutional capacity. A national AI policy that fails at citation verification symbolically demonstrates the very problem it seeks to regulate.

At the continental level, the African Union (AU) Data Policy Framework and Continental AI Strategy offer important guidance [4, 5]. The AU Data Policy Framework aims to strengthen and harmonise data governance frameworks and create a shared data space and standards for Africa. The Continental AI Strategy highlights the need for effective data governance systems, responsible AI, capacity building, and policy coordination. These instruments are valuable, but they are not directly enforceable South African law. They provide normative and policy guidance rather than binding obligations. South Africa therefore needs domestic legal and institutional arrangements that translate these principles into enforceable standards.

The overall legal picture is one of partial coverage. South Africa has constitutional rights, privacy laws, consumer laws, competition laws, administrative laws, cybercrime law, sectoral regulations, and emerging AI policy. What it does not yet have is a coherent AI governance framework capable of addressing the full data-colonial chain: extraction, storage, infrastructure, model development, deployment, auditing, value capture, public-interest benefit, contestation, and redress.

## 5. Analysis: Why Existing Laws do not Prevent Data Colonisation

The existing legal framework fails to prevent data colonisation for several reasons. We start by highlighting the issues that are inherent to South Africa's fragmented legal landscape in relation to AI. We then move on to discuss how the current legal framework is ineffective in addressing political power imbalances. Following this, we delve into the reasons why current laws are unable to prevent data colonialism from causing harm to individuals and society.

### 5.1. South Africa’s Fragmented Legal Landscape

South Africa’s current legal framework is fragmented across different statutes, each addressing particular rights, relationships or harms. South Africa's AI policy process has not yet become law. The withdrawal of the 2026 Draft National AI Policy does not mean South Africa has no plans for enacting AI specific policies. The 2024 Framework remains a useful starting point, and the policy revision process may lead to stronger governance. However, as of the current moment, there is no enacted AI regulation, no binding risk classification system, no AI regulator with statutory powers, and no binding national algorithmic accountability framework. This leaves South Africa dependent on existing laws that were not designed for the full political economy of AI.

Data colonisation operates structurally across sectors and institutions. POPIA gives rights to data subjects and imposes obligations on responsible parties in relation to the processing of personal information, including rights concerning access correction, objection and automated decision-making**.** PAIA (Promotion of Access to Information Act) facilitates access to records held by public bodies and, in specified circumstances, private bodies, while PAJA (Promotion of Administrative Justice Act) protects lawful, reasonable and procedurally fair administrative action [56, 57]**.** Consumer law addresses unfair market and consumer practices, while cybercrime and electronic-transactions legislation address other dimensions of the digital environment [58, 59, 61]**.** Yet no single component of this landscape is designed to address the cumulative concentration of informational, infrastructural, economic, and algorithmic power associated with data colonisation. Thus, accountability is also fragmented across regulators. The Information Regulator handles POPIA and access to information. The Competition Commission handles market dominance and competition. The FSCA and Prudential Authority handle financial-sector conduct and prudential risk. The Film and Publication Board, ICASA, SAPS, Home Affairs, SASSA, municipal authorities, and sector departments may all be implicated in different AI contexts. However, AI systems cut across these boundaries. A platform can raise consumer, labour, competition, privacy, tax, and data-sovereignty issues simultaneously. Without coordination, AI governance can fall between institutional mandates. UNESCO’s assessment similarly treats POPIA as an important foundation for AI governance, while identifying the need for a more comprehensive AI governance framework [70]. An AI governance framework can mitigate against harms which occur at group, community, or systemic level. A credit model may disadvantage a neighbourhood because of historical patterns of poverty. A recruitment model may reproduce racial, gender, disability, or language bias. A surveillance network may change the character of public space for everyone in a community. A recommender system may shape the visibility of local media, small businesses, or political content. In these contexts, the harm cannot be reduced to one person's privacy, one consumer transaction, or one administrative decision. W

This structural shortcoming manifests though several issues. South Africa lacks a public AI transparency infrastructure. There is no public register of high-risk AI systems used by government or powerful private actors. There is no standard requirement for algorithmic impact assessments before deployment in high-impact domains. There is no comprehensive public sector AI procurement standard requiring explainability, audit logs, data provenance, bias testing, human-review procedures, and public reporting. This means that many AI systems may be deployed without citizens knowing that automated tools are being used. Transparency cannot be left to private notices or voluntary ethics statements. Further, the South African legal framework falls short on enforcement capacity. The Information Regulator has an important mandate but faces the challenge of regulating a rapidly expanding data economy. AI governance requires legal knowledge, technical auditing capacity, data science expertise, procurement expertise, sectoral coordination, and investigative resources. Regulators must understand model documentation, training data, bias testing, data flows, cybersecurity, cloud contracts, and vendor architectures. Without this capacity, rights may remain formal. A regulator that cannot audit an AI system, cannot fully hold it accountable. Data colonisation thrives where local institutions lack the technical and financial resources to inspect the systems that govern them.

### 5.2. Power Imbalances

Existing laws regulate particular legal objects rather than the full process through which data become economic and governing power. POPIA regulates the processing of personal information, while AI systems can generate risk scores, behavioural predictions, similarity clusters, fraud flags, worker-performance metrics, neighbourhood classifications, or consumer segments from underlying data. These outputs may originate in personal information but can subsequently become proprietary scores, models or forms of derived intelligence. PAIA may provide access to identifiable records, but it is fundamentally an access-to-information statute rather than a framework allocating rights over algorithmically generated knowledge [57]. Consumer law may intervene when such intelligence produces an unfair consumer practice, but it does not determine who should control the economic value created from population-scale data [59]**.** This creates a major data-colonial gap: South Africans may generate the informational raw material, while firms control the profiles, models, scores, and predictive insights produced from it.

We also refer to this issue as value extraction. POPIA can require lawful processing, but it does not require benefit-sharing. It does not ask whether South African data is used to train profitable models whose benefits accrue elsewhere. It does not require AI firms to disclose the value generated from public datasets or communities. It does not create a data dividend, local reinvestment duty, or public-interest licensing obligation. Competition law can address some forms of platform dominance, but competition remedies are usually market-specific and reactive. Data colonisation requires a broader economic analysis: who captures value from South African data, and what mechanisms ensure that local communities, workers, creators, public institutions, and small businesses benefit?

(Political) power is also relevant when it comes to infrastructure and technological dependency. Data colonisation is partly constituted through cloud platforms, model-hosting environments, APIs, proprietary software, surveillance technologies, computing hardware and analytics vendors. POPIA regulates personal-information processing and cross-border data flows, while the Cybercrimes Act addresses cyber offences and ECTA regulates electronic communications and transactions. These protections are important, but they do not determine whether South Africa possesses sovereign capacity to inspect, alter or independently operate critical AI infrastructure. A public body or bank may satisfy its data-protection obligations while depending on a proprietary foreign platform for architecture, model APIs, updates, logs, security standards and computational capacity. Such dependence is not automatically unlawful and may provide significant technical advantages. The data-colonisation concern arises where dependency produces lock-in, limited auditability, restricted portability, weak bargaining power, or loss of public control over critical digital functions. The legal question therefore extends beyond whether personal information has been lawfully processed to who ultimately controls the infrastructures through which information becomes knowledge, prediction and decision-making.

Moreover, AI systems can be difficult to explain because they involve complex training data, probabilistic outputs, feature interactions, vendor secrecy, and layered deployment environments, which results in the problem of algorithmic opacity. Section 71 of POPIA provides important protections against certain decisions based solely on automated processing and requires safeguards in circumstances in which automated decision-making is permitted**.** However, South African scholarship has identified important limits in these protections. Novazi argues that POPIA does not establish an explicit and comprehensive right to an explanation [49], while other legal analysis similarly distinguishes the information about "underlying logic" contemplated by POPIA from a broader right to explanation [68]**.** PAJA provides rights to lawful, reasonable and procedurally fair administrative action and reasons in appropriate cases, but its focus is administrative action rather than algorithmic systems generally [56]. PAIA can facilitate access to existing records, but it does not create a general obligation on an AI provider to make models, training data, source code or proprietary decision architectures accessible to affected persons. Consumer, cybercrime and electronic-transactions legislation similarly do not establish a general system of algorithmic audits or explanations. This fragmentation matters because a right to object, request information or complain is weakened when people cannot determine how an algorithmic decision was reached. In data-colonial terms, opacity is itself a form of power.

### 5.3. Algorithmic Harms, Redress and Public Inclusion

Existing governance often treats AI as an innovation issue before treating it as a power issue. The national policy discourse emphasises economic growth, investment, skills, research, and competitiveness. But innovation without power analysis can reproduce colonial extraction. If local startups merely plug into foreign APIs, if public bodies outsource AI systems without audit rights, if local languages are underrepresented in models, if data centres are locally hosted but externally controlled, and if communities provide data without benefit-sharing, then AI innovation may deepen dependency. A decolonial approach does not reject innovation. It asks what kind of innovation, owned by whom, governed by whom, and for whose benefit.

The existing governance landscape provides forms of redress but does not necessarily ensure meaningful oversight of AI-mediated decisions. POPIA's section 71 focuses specifically on decisions based solely on automated processing**.** This creates an important governance question where a nominal human decision-maker is retained but relies heavily on an algorithmic recommendation. Evidence from research on automation bias shows that human decision-makers can selectively defer to algorithmic advice rather than independently scrutinising it [8]**.** Other legislation provides different routes to redress: PAJA permits challenges to qualifying administrative action, PAIA facilitates access to records, and the Consumer Protection Act provides consumer protections and enforcement mechanisms. But these mechanisms were not designed as an integrated system of AI oversight. A decolonial AI governance framework must therefore ask not simply whether a human appears somewhere in the decision chain or whether a legal remedy technically exists, but whether human review is meaningful, informed, independent, and empowered to reject the model output, and practically accessible to affected people.

However, public participation remains underdeveloped. While there is a plethora of public opinion research on AI from Western countries focusing, among others, on AI knowledge [12, 75], preference of ethical principles such as fairness, accountability, and transparency [38], risk acceptance of AI technologies [16, 65], or regulatory demands [39, 76], public opinion literature encompassing the South African context remains scarce (with a few exceptions such as [33, 53]). Yet, many implementation and design choices for AI systems are social choices, not purely technical decisions. South Africans should be involved in deciding how much risk is acceptable for AI systems used in policing, welfare, education, finance, healthcare, employment, and public space. Public participation is especially important in a society where technological harms may fall unevenly on poor, Black, rural, migrant, disabled, and digitally excluded communities. Data-colonial governance occurs when decisions about AI systems are made by vendors, experts, or officials without meaningful participation by those most affected.

Consequently, individual harm in the form of epistemic injustice is still a tangible effect of data colonisation. AI systems classify people and places through categories. If those categories are built from datasets that underrepresent African languages, informal economies, rural livelihoods, township mobility, historical exclusion, or local knowledge, then AI systems may misrecognise South African realities. Scholars have shown that algorithmic systems can reproduce racial and gendered harms when training data and design assumptions are biased [13, 17, 48]. In South Africa, such harms must be read against apartheid and colonial histories of classification. Data colonisation is therefore also epistemic: it concerns whose knowledge counts and whose realities are made machine-readable. Epistemic injustice is evident in many examples. Here, we focus on biometric surveillance and automated profiling.

In South Africa, where apartheid used identity, movement control, and surveillance as tools of racial governance, the issue of biometric surveillance is especially prevalent. Yet, Allen and van Zyl note that South Africa's data-protection laws do not fully cover the procurement and deployment of biometric surveillance hardware and AI tools [7]. POPIA treats biometric information as special personal information, but this does not amount to a dedicated biometric surveillance law. CCTV, licence-plate recognition, facial recognition, gait analysis, access-control systems, and behavioural analytics raise issues of public space, democratic freedom, protest, policing, racial bias, function creep, and private security power. However, South Africa lacks a comprehensive human-rights-based framework governing facial recognition and biometric surveillance [36, 69]. In comparison, the EU AI Act defines some forms of biometric surveillance, such as remote biometric identification (RBI) tools as either prohibited or high-risk tools, depending on the purpose of use. Either way, biometric surveillance tools in the European Union are under scrutiny and must be undergo elaborate risk assessment procedures before they can be used on the European market. This demand is also largely supported by the public as research on regulatory demands for RBI tools show [39].

Automated profiling is also central to AI. People can be profiled as risky borrowers, likely fraudsters, productive or unproductive workers, valuable customers, suspicious travellers, unreliable tenants, vulnerable consumers, or likely voters. However, Chivize argues that POPIA's current framework may be ill-equipped to mitigate automated decision-making risks in banking and financial inclusion. POPIA mentions profiling in section 71, but only in relation to solely automated decisions with significant effects [18]. It does not comprehensively regulate low-level profiling that accumulates over time or profiling that affects opportunities indirectly. Insufficient regulation causes risk as many AI harms are cumulative. A person may not experience a single dramatic automated denial, but may face higher prices, fewer opportunities, lower visibility, more scrutiny, or poorer service across multiple systems. Data colonisation operates through these accumulated asymmetries.

## 6. Towards Decolonial AI Governance in South Africa

Our analysis shows the insufficiency of current regulatory approaches to prevent data colonization in South Africa. South Africa's current laws regulate aspects of data processing, consumer harm, competition, cybersecurity, and administrative justice, but they do not prevent data colonisation. They address fragments of the problem while leaving the structural chain of extraction, infrastructure, modelling, value capture, opacity, and dependency largely intact – which results in tangible harms for individuals, especially marginalised communities.

In this section, we adapt a decolonial lens, i.e. we analyse pathways to a structural process of decolonialization that focuses on "power, economics, language, culture and thinking that shapes contemporary life" [45:664]. Taking this standpoint also "ensures that AI systems do not perpetuate historical injustices but instead contribute to a more balanced, inclusive, equitable, and just society." [74:256] A decolonial AI governance framework for South Africa should begin by accepting that data protection is necessary but insufficient. POPIA must be strengthened and enforced, but the country also needs tools that regulate AI systems as socio-technical infrastructures of power. The aim should not be to stop AI development, but to ensure that AI development supports constitutional democracy, equality, dignity, public accountability, local innovation, and shared economic benefit. In the following, we provide recommendations on how to do so.

South Africa should require algorithmic impact assessments for high-risk AI systems. The EU AI Act can serve as an inspirational starting point on how to define rules and classify systems; however, we argue that those definitions need to be adapted for the South African context and should not be copied as such. Impact assessments should be mandatory before deployment in areas such as policing, border control, welfare, healthcare, education, employment, credit, insurance, housing, municipal services, and biometric surveillance. They should assess data sources, purpose, affected groups, bias risks, explainability, human-review procedures, cybersecurity, procurement arrangements, cross-border data flows, model ownership, and redress mechanisms. Assessments should not be internal box-ticking exercises but also include the social context of system implementation. Individuals should receive clear and accessible notification when an AI or automated system is being used to make or materially influence a decision affecting them, including information about the system's purpose, the data used, the role of human oversight, and available mechanisms for explanation, review, and appeal. Public bodies should also disclose the deployment of such systems so that communities, researchers, and civil society can scrutinise their broader social effects.

South Africa should also create a public register of high-risk public-sector AI systems. Government departments, municipalities, state-owned entities, and public agencies should disclose where AI or automated decision-making is used, what the system does, who supplied it, what data it uses, whether it affects rights or benefits, how people can appeal, and when the system was audited. This would directly address the opacity issue. It would also help journalists, civil society, researchers, courts, and affected communities monitor public-sector automation.

Further, South Africa should adopt clear procurement rules for AI and data systems. Public bodies should not procure AI systems unless contracts guarantee audit rights, data-portability rights, documentation, access logs, bias testing, human-review procedures, cybersecurity standards, and restrictions on secondary data use. Vendors should not be allowed to use public-sector data to train proprietary models without explicit legal authorisation, public-interest safeguards, and benefit-sharing arrangements. Public procurement is one of the most powerful tools for preventing data colonisation because it can set the terms under which private vendors access public data [49].

POPIA should be updated or supplemented to address automated profiling, explanation, and collective harms. Section 71 should be clarified to require meaningful human review, not merely nominal human involvement. Affected persons should have a practical right to understandable reasons where AI contributes materially to a decision. Organisations deploying high-impact AI should be required to maintain model documentation, data provenance records, and audit trails. The law should also recognise that communities and groups can be harmed by profiling even where no individual can easily prove a privacy violation.

Moreover, South Africa should build regulatory capacity for algorithmic audits. The Information Regulator should have a dedicated AI and automated systems unit with legal, technical, and social-science expertise. This unit should coordinate with the Competition Commission, FSCA, Prudential Authority, ICASA, Human Rights Commission, Public Protector, Auditor-General, and sector regulators. It should be able to inspect model documentation, require independent audits, issue guidance, investigate complaints, and recommend legislative reform. Without technical capacity, even the strongest legal provisions will be difficult to enforce.

Further, South Africa should develop a public-interest data governance model. Not all data should be treated as private property or as a free resource for extraction. Public-sector data, community data, health data, language data, environmental data, and cultural data may require stewardship arrangements that balance innovation with democratic control. Data trusts, data commons, public-interest licences, community data governance protocols, and benefit-sharing models should be explored. The point is not to block data use, but to prevent one-way extraction.

Competition law should continue to address digital-market power and data-driven dominance. The Competition Commission's digital markets work should be linked more explicitly to AI governance. Dominant platforms should be required to provide fair access, transparent ranking, limits on self-preferencing, data-portability mechanisms, and remedies that support local media, small businesses, and historically disadvantaged market participants. The media and digital platforms inquiry is especially relevant because AI systems increasingly use news and cultural content as training or retrieval material. South Africa should ensure that creators, journalists, language communities, and media institutions are not reduced to unpaid data suppliers for global AI systems.

AI infrastructure policy should distinguish between data residency and digital sovereignty [63, 64]. Local data centres are useful, but they do not guarantee sovereignty if ownership, control, software architecture, chips, cloud services, and legal jurisdiction remain external. South Africa should develop a realistic AI infrastructure strategy that includes local compute capacity, public research infrastructure, open-source tools, skills development, interoperable standards, secure government cloud requirements, and vendor-diversification strategies. This should be done without technological nationalism for its own sake. The goal is practical regulatory and developmental agency and to achieve ownership of African data.

South Africa should also centre local languages, communities, and knowledge systems in AI development [1, 3, 47]. Data colonisation occurs when AI systems extract from local contexts without representing them fairly. South Africa's multilingual society requires AI systems that recognise linguistic diversity, township and rural realities, informal economies, disability access, and local knowledge. Additionally, African centred AI should also include African-specific values such as communitarianism and human dignity [11, 67] or respect for diversity and ethnic neutrality [44] as central building block for local technologies. Public funding for AI should support open, well-documented, rights-respecting datasets for South African languages and public-interest use cases, while protecting communities from exploitation.

Additionally, AI governance should be democratic [74]. Public consultations should not be limited to industry, government, and technical experts. Workers, unions, civil society, disability organisations, rural communities, township residents, consumer groups, migrants, students, women, and historically disadvantaged communities should be included as they can provide different views on AI impacts [15, 37]. Algorithmic systems affect lived realities. A decolonial AI framework must therefore allow those affected to shape the rules before harms occur [21, 26, 74].

Finally, South Africa's revised AI policy should be evidence-based, transparent, and legally operational. The withdrawal of the 2026 Draft AI Policy should be treated as an opportunity to produce a stronger document. The revised policy should clearly distinguish between principles, laws, institutions, enforcement powers, procurement standards, technical standards, and sector-specific rules. It should avoid vague ethics language unless tied to enforceable duties. It should also recognise data colonisation as a structural risk, not merely a privacy or innovation issue.

## 7. Conclusion

This article has argued that South Africa's AI governance challenge should be framed through data colonisation. The article's central argument is that South Africa's current laws do not adequately regulate AI-specific risks such as algorithmic opacity, foreign-controlled digital infrastructure, biometric surveillance, automated profiling, and unequal extraction of value from South African data. Data protection laws such as POPIA remains important, but they do not prevent data colonisation because data colonisation is not primarily about privacy. It is about ownership, dependency, surveillance, sovereignty, value extraction, and regulatory power. POPIA can regulate personal information, but it cannot by itself create algorithmic accountability, public-interest data governance, infrastructure sovereignty, or democratic control over AI systems.

A more adequate framework requires moving from privacy compliance to decolonial AI governance. This means mandatory algorithmic impact assessments, public-sector AI registers, independent audits, biometric safeguards, stronger rights to explanation and meaningful human review, procurement rules that protect public data, competition remedies against platform dominance, public-interest data stewardship, local AI infrastructure, and stronger regulatory capacity – all developed and maintained from a South African viewpoint. South Africa has the constitutional foundation and institutional resources to lead in this area. The question is whether it will regulate AI as a matter of public power and digital sovereignty, or whether it will allow the country's data to become another extractive frontier.

**Data availability statement.** Not applicable.

**Funding**: Kimon Kieslich was funded by the German Academic Exchange Service (DAAD) with funds from the Federal Ministry of Research, Technology and Space (BMFTR).

**Conflicts of interest/Competing interests**: The authors declare no competing interests.

**Ethical approval and informed consent statements**: Not applicable.

**Acknowledgement**: The authors would like to thank Kirsty Gibson-Kieslich for her substantive feedback on the legal analysis. Further, the authors thank the Centre for the Advancement of Scholarship (CAS) at the University of Pretoria for hosting both researchers and providing space for critical discussions.